\documentclass[sigconf]{acmart}



\usepackage{flafter}
\usepackage{placeins}
\usepackage{graphicx}
\usepackage{multirow}
\usepackage{booktabs}
\usepackage{enumitem}

\newcommand{\hdrup}{\rule{0pt}{2.6ex}}
\newcommand{\hdrdown}{\rule[-0.6ex]{0pt}{2.6ex}}
\usepackage{float}

\usepackage{algorithm}
\usepackage[noend]{algpseudocode}
\usepackage{amsmath}

\AtBeginDocument{%
  }

\title{ChainRec: An Agentic Recommender Learning to Route Tool Chains for Diverse and Evolving Interests}

\setcopyright{none}               
\renewcommand\footnotetextcopyrightpermission[1]{} 

\author{Fuchun Li}
\email{lifuchun25@ime.ac.cn}
\affiliation{%
  \institution{Chinese Academy of Sciences}
  \country{China}
}

\author{Qian Li}
\email{kathieqli@tencent.com}
\affiliation{%
  \institution{Tencent}
  \country{China}
}

\author{Xingyu Gao}
\email{gxy9910@gmail.com}
\affiliation{%
  \institution{Chinese Academy of Sciences}
  \country{China}
}

\author{Bocheng Pan}
\email{panbocheng@ime.ac.cn}
\affiliation{%
  \institution{Chinese Academy of Sciences}
  \country{China}
}

\author{Yang Wu}
\email{samuelywu@tencent.com}
\affiliation{%
  \institution{Tencent}
  \country{China}
}

\author{Jun Zhang}
\email{neoxzhang@tencent.com}
\affiliation{%
  \institution{Tencent}
  \country{China}
}

\author{Huan Yu}
\email{huanyu@tencent.com}
\affiliation{%
  \institution{Tencent}
  \country{China}
}

\author{Jie Jiang}
\email{zeus@tencent.com}
\affiliation{%
  \institution{Tencent}
  \country{China}
}

\author{Jinsheng Xiao}
\email{xiaojs@whu.edu.cn}
\affiliation{%
  \institution{Wuhan University}
  \country{China}
}

\author{Hailong Shi}
\email{shihailong2010@gmail.com}
\authornote{Corresponding author.}
\affiliation{%
  \institution{Chinese Academy of Sciences}
  \country{China}
}

\begin{abstract}
Large language models (LLMs) are increasingly integrated into recommender systems, motivating recent interest in agentic and reasoning-based recommendation. However, most existing approaches still rely on fixed workflows, applying the same reasoning procedure across diverse recommendation scenarios. In practice, user contexts vary substantially---for example, in cold-start settings or during interest shifts---so an agent should adaptively decide what evidence to gather next rather than following a scripted process. To address this, we propose ChainRec, an agentic recommender that uses a planner to dynamically select reasoning tools. ChainRec builds a standardized Tool Agent Library from expert trajectories. It then trains a planner using supervised fine-tuning and preference optimization to dynamically select tools, decide their order, and determine when to stop. Experiments on AgentRecBench across Amazon, Yelp, and Goodreads show that ChainRec consistently improves Avg HR@\{1,3,5\} over strong baselines, with especially notable gains in cold-start and evolving-interest scenarios. Ablation studies further validate the importance of tool standardization and preference-optimized planning.

\end{abstract}

\begin{document}
\sloppy

\maketitle


\section{Introduction}

Recommender systems are widely deployed across the Web in search, feeds, marketplaces, and advertising to rank and filter content and items. A central challenge is that user interests are diverse across domains and often evolve over time, making a single fixed recommendation strategy brittle across scenarios, such as cold-start users with sparse histories or interest-shift episodes where short-term intent conflicts with long-term tastes. A long line of prior work has developed strong temporal, sequential, and graph-based recommender models to address these challenges~\cite{Koren2009TimeSVD,Kang2018SASRec,Sun2019BERT4Rec,He2020LightGCN}. However, while powerful, these approaches are often designed around static input signals and fixed modeling pipelines, making them less flexible when user needs shift or when additional evidence must be actively gathered on demand. This motivates recommendation frameworks that can adapt their decision process dynamically by reasoning over available context and selectively acquiring new information, a setting where large language models (LLMs) are particularly well suited.

Agentic recommenders powered by LLMs offer a promising direction by enabling an agent to not only generate recommendations, but also reason, plan, and take actions to gather supporting evidence before ranking. Early studies show that chain-of-thought prompting can improve multi-step decision making by making intermediate reasoning explicit~\cite{Wei2022CoT}, and that tool use further equips agents to retrieve and process information through external functions~\cite{Yao2023ReAct,Schick2023Toolformer}. Building on these advances, recent work has explored LLM-based recommendation across a range of settings, including conversational preference elicitation, instruction tuning, planning, and tool-augmented recommendation agents~\cite{Geng2022P5,Hou2023InstructRec,Lin2023ChatRec,Wu2023LLM4RecSurvey,Zhang2023Agent4Rec,Wang2023RecMind}.

Yet most existing agentic recommenders still operate under an implicit assumption: the agent is provided with near-complete context and can follow a fixed reasoning script. In realistic interactive settings, however, the agent often starts with incomplete information and must actively decide what evidence to acquire next, such as whether to prioritize user-side preference signals or candidate-side review and metadata cues. In these cases, the key challenge shifts from simply producing a reasoning chain to conducting strategic evidence gathering and dynamic planning under changing scenarios.

To address this challenge, we propose \emph{ChainRec}, an agentic recommendation framework that decouples \emph{what} the system can do from \emph{how} it decides to do it. ChainRec introduces a reusable library of standardized reasoning tools for evidence acquisition and processing, while a learned planner dynamically routes among these tools at decision time instead of following a fixed script. The planner is trained in two stages: supervised fine-tuning (SFT) provides basic competence in tool use by imitating expert trajectories, and direct preference optimization (DPO) further improves routing decisions by favoring higher-utility tool chains under the same input~\cite{Ouyang2022InstructGPT,Rafailov2023DPO}. As a result, ChainRec can adaptively gather user-side or item-side evidence depending on the signals available in each scenario.

We evaluate ChainRec under the community’s interactive recommendation protocol using \emph{AgentRecBench}~\cite{Ge2025AgentRecBench,AgentSociety2025} across Amazon, Yelp, and Goodreads. In each episode, the agent is initialized with a target user and a fixed candidate set, and must acquire any additional evidence through tool calls before producing a ranking. Experimental results show that ChainRec consistently outperforms strong agentic and non-agentic baselines on Avg HR@\{1,3,5\}, with especially notable improvements in cold-start and evolving-interest scenarios. Ablation studies further confirm the contribution of each component, highlighting the benefits of standardized tools and preference-optimized dynamic planning.

\paragraph{Contributions.}
\begin{itemize}
  \item \textbf{Dynamic, scenario-aware planning for recommendation.}
  We propose \emph{ChainRec}, a dynamic agentic method that reframes recommendation as an online \emph{observe–decide–act} process with state-aware tool routing, replacing fixed, static workflows with instance-adaptive planning and termination.

  \item \textbf{Capability construction and strategic policy learning.}
  We automate tool construction by mining, clustering, and standardizing recurring LLM reasoning steps into a reusable \emph{Tool Agent Library} (unified I/O, structured memory writes), and train a planner with an SFT$\rightarrow$DPO recipe to learn \emph{state-aware} routing under simple feasibility masks.

  \item \textbf{Empirical validation across domains and scenarios.}
  On AgentRecBench (Amazon, Goodreads, Yelp), ChainRec improves Avg HR@\{1,3,5\} over strong agentic and non-agentic baselines, with notable gains in cold-start and interest-shift settings; ablations further isolate the benefits of the standardized capability layer and preference-optimized planning.
\end{itemize}

\section{Related Work}

\subsection{Chain-of-Thought (CoT) reasoning with LLMs}
Chain-of-Thought (CoT) prompting enables Large Language Models (LLMs) to articulate intermediate reasoning steps, improving both performance and interpretability on complex tasks~\cite{Wei2022CoT}. Zero-shot CoT elicits such behavior with simple instructions~\cite{Kojima2022ZeroShotCoT}; Self-Consistency and Least-to-Most enhance robustness and decomposition by sampling diverse chains and solving subproblems sequentially~\cite{Wang2023LeastToMost}. Beyond purely internal chains, ReAct interleaves reasoning with external actions~\cite{Yao2023ReAct}, while Toolformer shows LLMs can learn to call tools under weak supervision~\cite{Schick2023Toolformer}. Subsequent work explores structured search over reasoning traces, such as Tree-of-Thoughts and Graph-of-Thoughts for coordinated exploration~\cite{Yao2023ToT,Besta2023GoT}, as well as self-ask decomposition and plan-and-solve style prompting~\cite{Press2022SelfAsk,Wang2023PlanAndSolve}. Self-reflection mechanisms (e.g., Reflexion) further refine iterative decision making~\cite{Shinn2023Reflexion}.

In our setting, these advances motivate a capability-first view: CoT traces across domains repeatedly instantiate a compact set of core patterns. Rather than relying on ad hoc, text-only chains per episode, we mine expert CoT traces to discover, cluster, and standardize recurring steps (e.g., intent distillation, candidate-side evidence gathering) into a reusable Tool Agent Library with consistent I/O, enabling an agent to plan and route over named, callable capabilities.

\subsection{Agents for recommender systems}
Work using LLMs in recommendation has coalesced around two complementary lines: (i) agent-based user modeling (dialogue, memory, context gathering; e.g., InstructRec, Chat-Rec~\cite{Hou2023InstructRec,Lin2023ChatRec}), and (ii) agent-based recommendation (planning, tool use, evidence composition; e.g., RecMind~\cite{Wang2023RecMind} and multi-step planning such as Agent4Rec~\cite{Zhang2023Agent4Rec}). Earlier efforts that use LLMs as a unified interface to tasks and data (e.g., P$^5$~\cite{Geng2022P5}) complement these lines, and surveys synthesize how LLMs interact with collaborative signals~\cite{Wu2023LLM4RecSurvey}. Community resources—AgentSociety and AgentRecBench—establish interactive protocols and repeatedly expose the limits of static checklists (fixed CoT or fixed call orders)~\cite{AgentSociety2025,Ge2025AgentRecBench}. Beyond recommendation, general-purpose tool agents and lifelong planners (e.g., Gorilla for API calling, Voyager for skill acquisition) illustrate the value of standardized tool interfaces and iterative planning~\cite{Patil2023Gorilla,Wang2023Voyager}.

We make the capability layer explicit and standardized before policy learning, and then learn state-conditioned routing over these capabilities (SFT$\rightarrow$DPO). This differs from fixed-call-order pipelines and from single-agent long-prompt controllers that entangle reasoning and evidence, and it aligns the agent’s contract—acting to gather and synthesize evidence—with regularities observed in CoT traces. We next motivate why this capability-first, routing-second stance is necessary and sufficient for interactive recommendation.

\subsection{Fine-Tuning and Preference Alignment for LLM Agents}
\label{subsec:sft_rl_alignment}

Supervised fine-tuning (SFT).
Post-2023 work has shown that instruction-style SFT is an effective way to stabilize formats, reduce hallucination, and ground tool use for agentic tasks. Representative efforts include self-supervised tool calling with weak signals~\cite{Schick2023Toolformer}, lightweight instruction tuning pipelines (e.g., Alpaca) that improve adherence with modest resources~\cite{Taori2023Alpaca}, and open chat models that couple large-scale instruction data with safety tuning~\cite{Touvron2023Llama2}. In interactive recommendation, SFT serves to normalize the agent interface: capability I/O schemas, memory writes, and basic retrieval-orienting behaviors are made reliable before any policy-level optimization. API-grounded SFT that maps language to function signatures (e.g., Gorilla) provides a complementary signal for robust tool use~\cite{Patil2023Gorilla}.

\begin{figure*}[t]
  \centering
  \includegraphics[width=\textwidth]{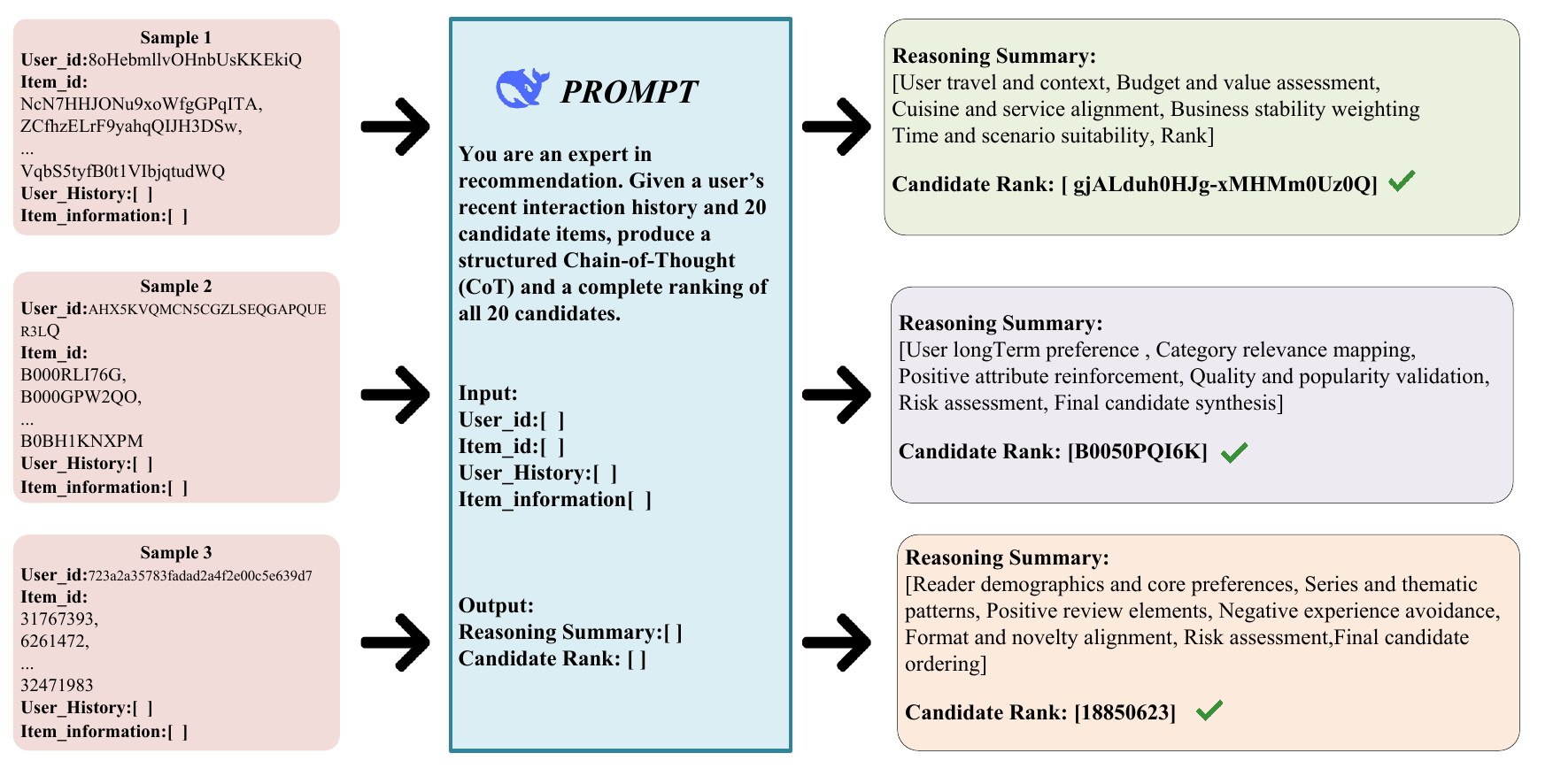}
 \caption{Scenario-dependent CoT under the same prompt. Three samples yield distinct reasoning routes yet valid rankings, motivating dynamic planning that decomposes steps and learns to route tools by state.}
  \Description{An overview with three horizontal panels for e-commerce, books, and local services, showing stepwise agent workflows from preference analysis through evidence gathering to final ranking.}
  \label{fig:teaser}
\end{figure*}

\begin{figure}[!htbp]
  \centering
  \includegraphics[width=\columnwidth]{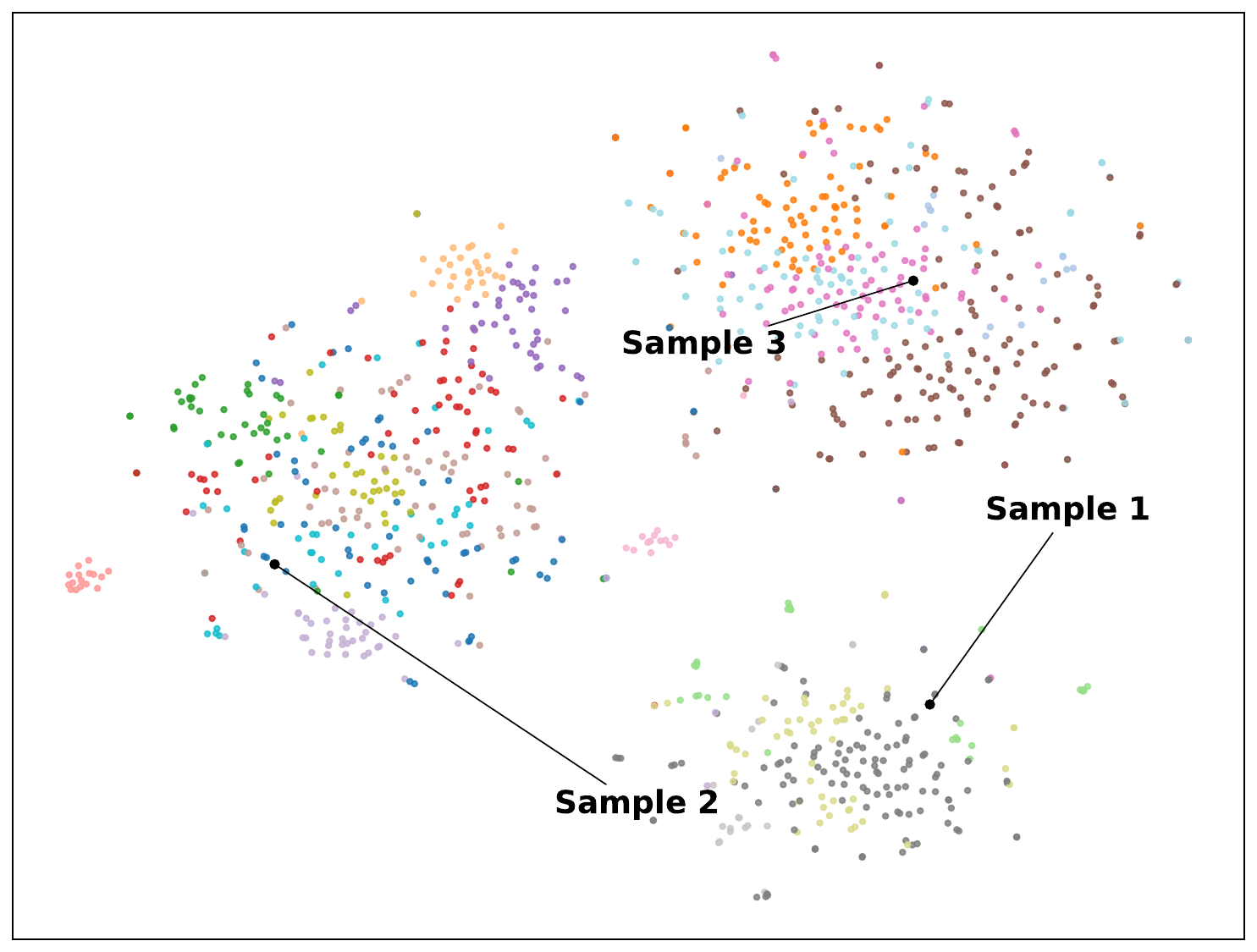}
  \caption{t-SNE of full CoT embeddings under the same prompt. Colors show clusters; Sample 1–3 mark three runs from different scenarios, illustrating scenario-induced CoT differences.}
  \label{fig:tsne}
\end{figure}

Preference and RL alignment.
Recent alignment methods move beyond reward-model–centric RLHF to preference objectives that operate directly on paired or ranked outputs. Direct Preference Optimization (DPO) optimizes a closed-form objective derived from pairwise preferences and works fully offline, avoiding reward modeling and on-policy rollouts~\cite{Rafailov2023DPO}. RRHF learns from listwise rankings to discourage suboptimal responses~\cite{Liu2023RRHF}. RLAIF replaces human labels with AI feedback to scale preference data~\cite{Lee2023RLAIF}. Follow-ups such as ORPO, SimPO, and KTO extend the objective design for stability, simplicity, or risk sensitivity~\cite{Song2024ORPO,Lu2024SimPO,Ethayarajh2024KTO}. These approaches are well-suited to agent settings where high-quality trajectories or comparisons can be collected offline and where environment interactions are expensive or rate-limited.

We adopt a two-stage recipe that is standard in recent LLM agents but tailored to interactive recommendation: SFT first ensures tool I/O and memory operations are dependable; preference optimization then aligns the planner for state-conditioned routing under a fixed candidate protocol and a small step budget. Compared with long-prompt controllers or fixed call orders, this separation keeps capability execution reliable while letting the policy learn which capability to invoke and when. In our setting, DPO is a practical choice because it trains from offline comparisons of tool-call sequences and improves routing quality without requiring longer plans.


\section{Motivation and Problem Definition}
\label{sec:sec3}

\subsection{Why dynamic planning is needed in recommendation scenarios}
\label{sec:motivation}

We conduct a large number of recommendation trials using the same prompting template across multiple real-world scenarios. Although the prompt remains identical, we find that the model often follows different reasoning routes depending on the scenario. Figure~\ref{fig:teaser} illustrates three representative examples (Sample~1--3), where the model focuses on different types of evidence before producing a ranking. Figure~\ref{fig:tsne} further confirms this phenomenon at scale: embedded reasoning traces form multiple distinct clusters rather than collapsing into a single uniform pattern.

These results highlight a simple but important point: different recommendation scenarios require different information. As a result, a fixed workflow with a prescribed tool order can be brittle. For instance, in cold-start settings with little user history, spending steps on long-term preference reasoning may provide limited benefit. Under interest shifts, over-relying on past behavior can miss the user’s immediate intent. When item-side signals are sparse or noisy, failing to retrieve supporting evidence can lead to poorly grounded rankings. Therefore, recommendation agents should plan dynamically at inference time---deciding what evidence to gather next, how to combine it, and when to stop. This motivates the sequential decision formulation introduced in the next section.



\subsection{Problem Definition}
\label{subsec:problem}

\paragraph{Task.}
Given a target user $u\in\mathcal{U}$ and a set of candidate items $I_{\text{cand}}$, the system must output a ranked list $L_{\text{ranked}}$ over $I_{\text{cand}}$. The environment contains user histories and item/review corpora, but this information is \emph{not} revealed upfront. Instead, the agent must actively acquire and summarize any needed evidence through tool calls before using it for ranking. This setting follows the interactive protocol of AgentRecBench.

\paragraph{Evaluation.}
We evaluate ranking quality using Hit Rate (HR) at cutoffs $K\in\{1,3,5\}$:
\begin{equation}
\mathrm{HR@}K \;=\; \mathbb{I}\!\big(\mathrm{rank}(i_{\text{pos}})\le K\big),
\end{equation}
where $i_{\text{pos}}$ is the held-out positive item and $\mathbb{I}(\cdot)$ is the indicator function. Following the benchmark, we report Avg HR@\{1,3,5\} as the main metric.

\paragraph{Decision process.}
To capture the agent’s sequential evidence-gathering behavior, we model the recommendation procedure as a finite-horizon Markov Decision Process (MDP)
$\mathcal{M}=(\mathcal{S},\mathcal{A},P,R)$ with a sparse terminal reward ($\gamma=1$).

\begin{itemize}[leftmargin=*]
  \item \textbf{State $\mathcal{S}$.}
  At step $t$, the observable state is
  \begin{equation}
  S_t \;=\; \big(u,\; I_{\mathrm{cand}},\; M_t\big),
  \end{equation}
  where $M_t=[(a_0,o_0),\ldots,(a_{t-1},o_{t-1})]$ stores past actions and their summarized outputs.
  Raw histories or reviews are not included unless they are explicitly acquired and written into memory. Thus, the state evolves as the agent collects more evidence over time.

  \item \textbf{Actions $\mathcal{A}$.}
  The agent can choose from a finite set of evidence-related tool operations $\{T_1,\ldots,T_K\}$,
  as well as a terminal ranking action $a_{\text{term}}$:
  \begin{equation}
  \mathcal{A} \;=\; \{T_1,\ldots,T_K\}\,\cup\,\{a_{\text{term}}\}.
  \end{equation}
  Each $T_k$ corresponds to a tool that retrieves or processes user/item evidence needed for recommendation.

  \item \textbf{Transition $P$.}
  Executing action $a_t$ produces a summarized output $o_t$, which is deterministically appended to memory:
  \begin{equation}
  S_{t+1}=\delta(S_t,a_t), 
  \qquad
  M_{t+1}=M_t \oplus [(a_t,o_t)].
  \end{equation}

  \item \textbf{Policy $\pi$.}
  A parameterized policy $\pi_\theta(a_t\mid S_t)$ selects the next feasible action,
  subject to simple constraints such as tool preconditions and a maximum step budget $T_{\max}$.

  \item \textbf{Reward $R$.}
  The agent receives a terminal utility that balances ranking quality and planning cost:
  \begin{equation}
  R(\tau)=\mathrm{Quality}\!\big(L_{\mathrm{ranked}}\big)-\lambda\,|\tau|,
  \end{equation}
  where $\mathrm{Quality}$ is instantiated as Avg HR@\{1,3,5\},
  $|\tau|$ counts the number of non-terminal tool steps,
  and $\lambda\ge 0$ controls the trade-off between accuracy and cost.
\end{itemize}

\noindent\textbf{Objective.}
Our goal is to learn an optimal policy
\begin{equation}
\pi^{*}=\arg\max_{\pi}\;\mathbb{E}_{\tau\sim p_{\pi}(\tau)}\big[R(\tau)\big],
\end{equation}
i.e., a policy that maximizes ranking performance while using as few tool steps as possible under the interaction protocol and step budget.

\begin{figure*}[t]
  \centering
  \includegraphics[
    width=\textwidth,
    trim=2pt 2pt 2pt 2pt,
    clip
  ]{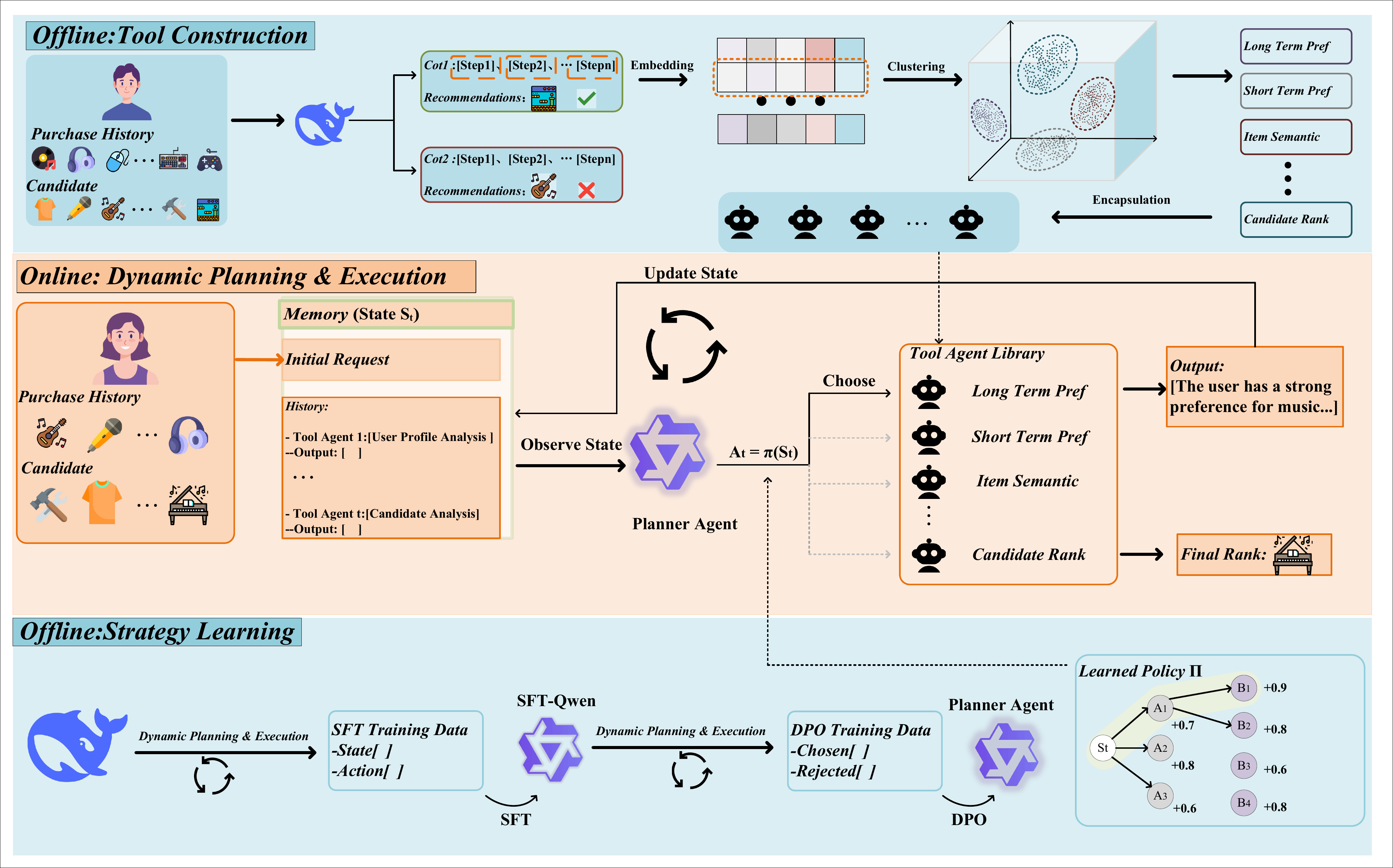}
  \caption{ChainRec offline--online workflow. Offline: capability construction by mining expert CoT traces and encapsulating them into tools; planner training with SFT $\rightarrow$ DPO. Online: the Planner observes the current state in Memory, selects a tool from the Tool Agent Library, executes it to obtain evidence, updates Memory, and repeats until CandidateRank outputs the final ranking.}
  \label{fig:framework}
\end{figure*}


\section{Method: ChainRec}
\label{sec:method}

\noindent\textbf{Overview.}\;
ChainRec replaces fixed recommendation pipelines with \emph{dynamic evidence gathering and planning}. 
Intuitively, instead of following a scripted tool order, the agent repeatedly decides what information to collect next before producing the final ranking.
The system consists of two layers:
(i) a \emph{tool layer} that provides reusable evidence-gathering and processing operations, forming a standardized \emph{Tool Agent Library (TAL)}; and
(ii) a \emph{policy layer} that trains a \emph{Planner} to select and compose tools based on the evolving state and accumulated memory.
To mirror actual usage, we describe the method in execution order:
we first present \S\ref{subsec:inference_first} \emph{online inference and dynamic execution},
then \S\ref{subsec:tools} \emph{offline tool construction},
and finally \S\ref{subsec:learning} \emph{offline strategic learning}.

\subsection{Online Inference and Dynamic Execution}
\label{subsec:inference_first}

\begin{algorithm}[!t]
  \caption{ChainRec: Online Dynamic Execution}
  \label{alg:chainrec_inference}
  \begin{algorithmic}[1]
    \Require target user $u_{\mathrm{target}}$; candidates $I_{\text{cand}}$; planner policy $\pi_{\theta}$; Tool Agent Library $\mathrm{TAL}$
    \Ensure final ranked list $L_{\mathrm{ranked}}$
    \State $M \gets \textsc{InitializeMemory}(u_{\mathrm{target}}, I_{\text{cand}})$;\; $S_0 \gets \textsc{GetState}(M)$;\; $t \gets 0$
    \While{true}
      \State $\mathcal{A}_{\mathrm{feasible}} \gets \textsc{FilterTools}(\mathrm{TAL}, S_t)$ 
      \Comment{prevent invalid calls, no repeats, step budget}
      \State $A_t \gets \arg\max_{a \in \mathcal{A}_{\mathrm{feasible}}} \pi_{\theta}(a \mid S_t)$ 
      \Comment{Planner selects next tool}
      \State $O_t \gets \textsc{ExecuteTool}(A_t, S_t)$ 
      \Comment{returns a short structured evidence summary}
      \State $M \gets \textsc{UpdateMemory}(M, A_t, O_t)$;\; $S_{t+1} \gets \textsc{GetState}(M)$;\; $t \gets t+1$
      \If{$A_t = a_{\text{terminate}}$}
        \State $L_{\mathrm{ranked}} \gets O_t$ 
        \Comment{\texttt{CandidateRank} emits the final ranking}
        \State \textbf{break}
      \EndIf
    \EndWhile
    \State \Return $L_{\mathrm{ranked}}$
  \end{algorithmic}
\end{algorithm}

At inference time, ChainRec follows an \emph{observe--decide--act} loop that incrementally gathers evidence until termination.
The loop is coordinated by three components:
the \textbf{Planner} (decision maker), the \textbf{TAL} (available tool actions), and \textbf{Memory} (accumulated evidence summaries).

\paragraph{State composition and initialization.}
Given a task instance, we initialize Memory and form the initial state
\begin{equation}
S_0=(u,\; I_{\text{cand}},\; M_0),
\end{equation}
where $u$ is the target user description, $I_{\text{cand}}$ the candidate set, and 
$M_t=[(a_0,o_0),\ldots,(a_{t-1},o_{t-1})]$ records past tool calls and their summarized outputs.
Raw histories or reviews are not directly included unless a tool explicitly retrieves and summarizes them into $M_t$.
Thus, the state evolves as the agent collects more evidence over time.

\paragraph{Feasibility constraints.}
At step $t$, the Planner selects only from a feasible action set.
These constraints simply prevent invalid tool calls
(e.g., calling ranking before collecting any evidence), avoid degenerate loops, 
and enforce a maximum step budget $T_{\max}$.

\paragraph{Selection and execution.}
The Planner scores feasible actions and selects $A_t$.
Executing tool $A_t$ follows a fixed interface and returns a structured evidence summary $O_t$.
We then append $(A_t,O_t)$ into Memory and update the state deterministically:
\begin{equation}
S_{t+1}=\delta(S_t,A_t), \qquad M_{t+1}=M_t \oplus [(A_t,O_t)].
\end{equation}
This shared writing schema simplifies downstream planning decisions.

\paragraph{Termination and ranking.}
When the Planner selects the terminal action $a_{\text{terminate}}$ (\texttt{CandidateRank}), 
the system outputs the final ranked list $L_{\text{ranked}}$ and halts.
Because the Planner re-evaluates actions after every memory update, evidence acquisition is instance-adaptive rather than scripted.

\noindent\textit{Discussion.}\;
The online routine captures the intended contract: the agent decides \emph{what} evidence to acquire, 
\emph{in what order}, and \emph{when} to stop, based on the evolving state.
Next, we explain how we build the standardized tool layer and train a reliable Planner.

\subsection{Offline Tool Construction}
\label{subsec:tools}

We derive a compact set of reusable tools from expert-level chain-of-thought (CoT) traces in three stages.

\paragraph{(i) Collecting reasoning chains.}
Using a unified prompting template aligned with the evaluation protocol, a strong reasoning LLM produces step-labeled CoT traces and rankings for training tasks.
We retain chains that are both correct and concise (e.g., HR@5$=1$, within step budget $T\le T_{\max}$, and no immediate repeats).
When multiple valid chains exist, we apply self-consistency and keep the shorter majority outcome, yielding a clean corpus $\mathcal{D}_{\mathrm{CoT}}$.

\paragraph{(ii) Step normalization and clustering.}
Our goal is to identify recurring reasoning steps across expert traces, so that each can be turned into a reusable tool.
From each chain we extract step descriptors and normalize them into $(op,\; args)$ under a finite action vocabulary.
We embed each step into an L2-normalized vector $\mathbf{e}(r)\in\mathbb{R}^d$ and cluster with $k$-means by minimizing
\begin{equation}
\label{eq:kmeans}
\min_{\{\mathbf{c}_j\}_{j=1}^{k}} \; J(k) \;=\; \sum_{j=1}^{k} \sum_{r\in\mathcal{C}_j} \bigl\|\mathbf{e}(r)-\mathbf{c}_j\bigr\|_2^2.
\end{equation}
We select $k$ via an elbow on $J(k)$ and a silhouette plateau.
Clustering is performed per domain and consolidated under shared tool names.

\paragraph{(iii) Encapsulation into tools.}
Each cluster is encapsulated as a callable tool with a fixed name, input format, and structured output:
\[
\textit{Tool}=\langle name,\; desc,\; inputs,\; outputs\rangle.
\]
Outputs follow a compact schema
\[
\textit{outputs}=\{\textit{facets}:[(k_i,v_i)],\; \textit{confidence}\in[0,1]\}.
\]
This standardization ensures that evidence is written into Memory in a consistent way.
The resulting Tool Agent Library,
\[
\mathit{TAL}=\{Tool_1,\ldots,Tool_K\},
\]
provides the discrete action space for planning.
Implementation uses lightweight prompts with consistent I/O; examples and the full catalog appear in the appendix.

\subsection{Offline Strategic Learning}
\label{subsec:learning}

With TAL ready, we train a Planner to learn a dynamic policy rather than a fixed call order.

\paragraph{Trajectory data.}
We constrain a strong LLM to solve training tasks using only tools from TAL.
Successful solutions yield tool-call sequences together with intermediate summaries, forming trajectories for policy learning.

\paragraph{Supervised fine-tuning (SFT).}
We cast planning as behavior cloning on expert trajectories.
Each trajectory $\tau=(s_0,a_0,\dots,s_T,a_T)$ expands into $(s_t,a_t)$ pairs, where $s_t$ includes the task specification and accumulated Memory up to step $t$.
The SFT loss is the masked cross-entropy:
\begin{equation}
\mathcal{L}_{\mathrm{SFT}}
= -\sum_{t=0}^{T}\log p_\theta\!\big(a_t\mid s_t\big).
\end{equation}

\paragraph{Direct Preference Optimization (DPO).}
SFT teaches how to imitate expert tool choices, while DPO further teaches which tool chains to prefer.
Under a step budget $T$, we score a trajectory by
\begin{equation}
R(\tau)=\text{Quality}(L_{\text{ranked}})-\lambda\,|\tau|,
\end{equation}
where $\text{Quality}$ is Avg HR@\{1,3,5\} and $|\tau|$ counts tool invocations.
We sample diverse trajectories from the SFT policy and form preference pairs, where ties favor fewer steps.
DPO then increases the likelihood of preferred trajectories while keeping the policy close to the SFT reference.
After DPO, the Planner composes tools with improved overall utility, yielding higher ranking quality with comparable plan lengths.

\medskip\noindent\textbf{Summary.}\;
Execution-time adaptivity (\S\ref{subsec:inference_first}) is enabled by 
(i) a standardized tool layer (\S\ref{subsec:tools}) and 
(ii) a Planner trained for state-aware routing decisions (\S\ref{subsec:learning}).
This ordering mirrors actual usage: act dynamically at test time, supported by offline tool construction and strategic learning.

\section{Experiment}

\subsection{Experimental Setup}

\medskip
\noindent\textbf{Interactive Environment.}
We evaluate under the official interactive protocol of \emph{AgentRecBench}.
Each episode is initialized with only the target user and a fixed candidate set:
\[
O_0=(u_{\mathrm{target}},\, I_{\mathrm{cand}}), \qquad |I_{\mathrm{cand}}|=20.
\]

Following the benchmark construction, the candidate set contains exactly one held-out
ground-truth positive item $i_{\text{pos}}$ (a true future interaction of the user)
and 19 negative items sampled from the user’s unobserved (non-interacted) items
within the same domain.
Thus, each episode corresponds to a 1-positive/19-negative ranking task over a
fixed-size candidate pool.

All user histories, item metadata, and review corpora reside in the environment but are
\emph{not} revealed upfront.
Instead, the agent must actively retrieve any needed evidence via tool calls and write
summarized results into memory.

Online inference follows the observe--decide--act loop in Alg.~\ref{alg:chainrec_inference}:
at step $t$ the agent observes $S_t=(u, I_{\mathrm{cand}}, M_t)$, selects a feasible tool,
executes it, and appends $(a_t,o_t)$ into $M_{t+1}$ until the terminal action
\texttt{CandidateRank} outputs a ranking over $I_{\mathrm{cand}}$.
We report Avg HR@\{1,3,5\} under the official evaluation protocol.

\medskip
\noindent\textbf{Datasets and Scenarios.}
Experiments are conducted on the official AgentRecBench benchmark across three domains:
Amazon, Goodreads, and Yelp.

AgentRecBench instantiates each domain using curated real-world subsets.
For Amazon, we follow the benchmark setting built from three product subdomains:
\textit{Industrial\_and\_Scientific}, \textit{Musical\_Instruments}, and \textit{Video\_Games}.
For Goodreads, the environment is constructed from three book subdomains:
\textit{Children}, \textit{Comics\&Graphic}, and \textit{Poetry}.
For Yelp, we adopt the official Yelp Open Dataset provided by the benchmark.

Within each domain, AgentRecBench defines five scenario types:
\emph{Classic}, \emph{Cold-Start} (User/Item), and \emph{Evolving-Interest} (Long/Short).
Classic tasks correspond to standard recommendation settings with sufficient historical signals.
Cold-start tasks evaluate generalization under sparse interaction histories, where either the
target user has very limited observed interactions (User Cold-Start) or the positive item comes
from the cold-item subset (Item Cold-Start).
Evolving-interest tasks assess adaptability to dynamic user preferences via temporally segmented
histories, including a three-month interaction window (Long) and a one-week recent window (Short).
We follow AgentRecBench’s evolving-interest protocol, which simulates preference shifts
through these temporally segmented interaction windows.

\begin{table*}[t]
  \centering
  \caption{Overall results (Avg.\ HR@\{1,3,5\}) organized as Domain $\rightarrow$ Scenario (Classic / Cold-Start / Evolving-Interest) $\rightarrow$ Split (User/Item or Long/Short). “$\Delta$ vs best” denotes ChainRec minus the best Agentic Recommender Systems baseline in each column.}

\label{tab:all_results_domain_first}
\small
\setlength{\tabcolsep}{1.5pt}
\renewcommand{\arraystretch}{1.10}
\begin{tabular*}{\textwidth}{@{\extracolsep{\fill}} l l ccccc ccccc ccccc}
  \toprule
  \multicolumn{2}{c}{} &
  \multicolumn{5}{c}{Amazon\hdrup} &
  \multicolumn{5}{c}{Goodreads\hdrup} &
  \multicolumn{5}{c}{Yelp\hdrup} \\
  \cmidrule(lr){3-7}\cmidrule(lr){8-12}\cmidrule(lr){13-17}
  \multicolumn{2}{c}{} &
  Classic\hdrup & \multicolumn{2}{c}{Cold-Start\hdrup} & \multicolumn{2}{c}{Evo-Int\hdrup} &
  Classic\hdrup & \multicolumn{2}{c}{Cold-Start\hdrup} & \multicolumn{2}{c}{Evo-Int\hdrup} &
  Classic\hdrup & \multicolumn{2}{c}{Cold-Start\hdrup} & \multicolumn{2}{c}{Evo-Int\hdrup} \\
  \cmidrule(lr){3-3}\cmidrule(lr){4-5}\cmidrule(lr){6-7}
  \cmidrule(lr){8-8}\cmidrule(lr){9-10}\cmidrule(lr){11-12}
  \cmidrule(lr){13-13}\cmidrule(lr){14-15}\cmidrule(lr){16-17}
  Category & Method &
       & \makebox[1.8em][c]{User\hdrdown} & \makebox[1.8em][c]{Item\hdrdown}
       & \makebox[1.9em][c]{Long\hdrdown} & \makebox[1.9em][c]{Short\hdrdown} &
       & \makebox[1.8em][c]{User\hdrdown} & \makebox[1.8em][c]{Item\hdrdown}
       & \makebox[1.9em][c]{Long\hdrdown} & \makebox[1.9em][c]{Short\hdrdown} &
       & \makebox[1.8em][c]{User\hdrdown} & \makebox[1.8em][c]{Item\hdrdown}
       & \makebox[1.9em][c]{Long\hdrdown} & \makebox[1.9em][c]{Short\hdrdown} \\
  \midrule
  Traditional RS & MF
  & 15.0 & 15.0 & 15.0 & 38.4 & 49.5
  & 15.0 & 15.0 & 15.0 & 29.7 & 21.3
  & 15.0 & 15.0 & 15.0 & 42.0 & 65.9 \\
  DL-based RS & LightGCN
  & 15.0 & 15.0 & 15.0 & 32.3 & 59.1
  & 15.0 & 15.0 & 15.0 & 18.0 & 22.3
  & 15.0 & 15.0 & 15.0 & 31.6 & 68.9 \\
  \midrule

  \multirow{8}{*}{Agentic RS} & BaseAgent
  & 44.0 & 16.3 & 14.0  & 17.3 & 19.7
  & 20.3 & 22.3 & 16.2 & 32.0 & 29.7
  & 4.0  & 4.0  & 4.0  & 5.7  & 4.3 \\
  & CoTAgent
  & 39.7 & 19.3 & 9.0 & 17.7 & 16.0
  & 23.0 & 20.3 & 14.5 & 24.0 & 16.3
  & 4.3  & \textbf{4.3}  & \textbf{4.3}  & 3.3  & 3.0 \\
  & MemoryAgent
  & 43.7 & 15.3 & 11.0 & 18.3 & 17.7
  & 18.0 & 16.7 & 15.5 & 29.3 & 29.7
  & 3.7  & \textbf{4.3}  & \textbf{4.3}  & 5.0  & 4.0 \\
  & CoTMemAgent
  & 33.3 & 16.7 & 12.0  & 13.3 & 18.7
  & 17.3 & 16.7 & 15.5 & 23.3 & 21.3
  & 4.3  & 3.7  & \textbf{4.3}  & 4.3  & 4.0 \\
  & Baseline666
  & 60.0 & 50.3 & \textbf{48.7} & 50.7 & \textbf{71.3}
  & 54.7 & 38.7 & \textbf{49.5} & 66.0 & 63.3
  & 7.3  & 1.3  & 2.7  & 0.0  & 0.0 \\
  & DummyAgent
  & 54.0 & 59.0 & 45.0 & \textbf{65.0} & 65.7
  & \textbf{56.3} & 37.7 & \textbf{49.5} & 66.7 & 60.7
  & 6.3  & 1.3  & 2.7  & \textbf{10.3}  & 6.3 \\
  & RecHacker
  & \textbf{63.0} & \textbf{59.7} & 47.0 & 64.3 & 68.0
  & 55.0 & \textbf{49.3} & 46.1 & \textbf{68.7} & \textbf{66.3}
  & 7.0  & 3.0  & 3.3  & 9.3  & \textbf{7.7} \\
  & Agent4Rec
  & 28.3 & 45.6 & 28.0 & 34.0 & 46.3
  & 9.3  & 37.3 &  11.1 & 41.3 & 42.7
  & \textbf{7.6}  & 2.7  & 0.7  & 10.0  & 6.0 \\
  \midrule

  \textbf{Ours} & \textbf{ChainRec}
  & \textbf{72.3} & \textbf{70.3} & \textbf{60.0} & \textbf{68.7} & \textbf{74.3}
  & \textbf{56.7} & \textbf{67.7} & \textbf{50.5} & \textbf{70.3} & \textbf{69.7}
  & \textbf{28.3} & \textbf{13.7} & \textbf{13.7} & \textbf{15.7} & \textbf{14.0} \\
  & \textit{Improve vs best }
  & {\footnotesize\itshape $+14.76\%$} & {\footnotesize\itshape $+17.75\%$} & {\footnotesize\itshape $+23.20\%$} & {\footnotesize\itshape $+5.69\%$} & {\footnotesize\itshape $+4.21\%$}
  & {\footnotesize\itshape $+0.71\%$}  & {\footnotesize\itshape $+37.32\%$} & {\footnotesize\itshape $+2.02\%$}  & {\footnotesize\itshape $+2.33\%$} & {\footnotesize\itshape $+5.13\%$}
  & {\footnotesize\itshape $+272.37\%$} & {\footnotesize\itshape $+218.60\%$} & {\footnotesize\itshape $+218.60\%$} & {\footnotesize\itshape $+52.43\%$} & {\footnotesize\itshape $+81.82\%$} \\

  \bottomrule
\end{tabular*}
\end{table*}

Beyond the official evaluation splits, we construct a 4{,}500-task training corpus aligned with
the same domains and scenario definitions for tool construction and policy learning.
We enforce strict de-duplication against all test user--item pairs and do not access any
instance-level texts or labels from the test split, preventing information leakage.
Full preprocessing and scenario construction details are provided in the appendix.

\medskip
\noindent\textbf{Implementation Details.}
\textit{Tool Agent Library (TAL).} The automated tool construction pipeline is described in Sec.~4.2, and the complete tool catalog with unified I/O schemas is provided in the appendix. To ensure a fair comparison with prior agentic baselines on AgentRecBench, we strictly follow the official benchmark configuration for tool execution: all tool agents are instantiated with the same backbone LLMs used by baseline systems, including \textbf{Qwen2.5-72B-Instruct}~\cite{Qwen2024TechReport} and \textbf{DeepSeek-V3}~\cite{DeepseekV3TR}. In contrast, the \emph{Planner} is implemented with a smaller model, \textbf{Qwen3-8B}~\cite{Yang2025Qwen3}, further post-trained via our SFT$\rightarrow$DPO pipeline to learn dynamic tool routing decisions. Following the benchmark protocol, we set the tool-agent temperature to 0 in all runs for reliability.

Since the official implementations of several agentic baselines in AgentRecBench are not publicly released, we directly cite their reported results from the benchmark paper. All comparisons are conducted under the same evaluation protocol and matched backbone configurations, ensuring fairness.

\medskip
\noindent\textbf{Planner, backbones, and training.}
ChainRec uses a learned planner policy $\pi_{\theta}$ trained with SFT$\rightarrow$DPO. In ablations we compare (a) an SFT-only planner (without DPO), (b) DeepSeek-R1 as the planner~\cite{DeepSeekR1}, and (c) the full ChainRec planner. All planners share identical prompts, the same Tool Agent Library (TAL), and the same CandidateRank prompt to isolate the effect of routing. Training details: backbone Qwen/Qwen3-8B with QLoRA (4-bit NF4, bf16, gradient checkpointing); optimizer paged\_adamw\_8bit. SFT: max\_len=512, 3 epochs, lr $2\times 10^{-5}$, per-device batch=4, grad-accum=8. DPO: continue from the SFT QLoRA adapter, max\_len=2048, 1 epoch, lr $1\times 10^{-5}$, $\beta=0.1$, per-device batch=2, grad-accum=16, warmup\_steps=50.

\subsection{Results and Analysis}

\paragraph{Overall findings and key observations.}
Across Amazon, Goodreads, and Yelp—under Classic, Cold-Start, and Evolving-Interest settings—ChainRec consistently surpasses the strongest agentic baseline in every column (Table~\ref{tab:all_results_domain_first}).
Representative margins ($\Delta$ vs.\ best) are:
Amazon—Classic $+14.76\%$, CS-User $+17.75\%$, CS-Item $+23.20\%$, Evo-Long $+5.69\%$, Evo-Short $+4.21\%$;
Goodreads—Classic $+0.71\%$, CS-User $+37.32\%$, CS-Item $+2.02\%$, Evo-Long $+2.33\%$, Evo-Short $+5.13\%$;
Yelp—Classic $+272.37\%$, CS-User $+218.60\%$, CS-Item $+218.60\%$, Evo-Long $+52.43\%$, Evo-Short $+81.82\%$.

\paragraph{Classic.}
Across domains, ChainRec surpasses the strongest agentic baseline in Classic by $+14.76\%$ on Amazon, $+0.71\%$ on Goodreads, and $+272.37\%$ on Yelp (Table~\ref{tab:all_results_domain_first}). 
Gains come from a state-aware mix of ShortTermPreference/LongTermPreference (user-side) with ItemSemantic/ItemProfile (candidate-side), all written in structured form to memory and consumed by CandidateRank. 
Where signals are mixed, PositivePreference/NegativePreference help separate what to favor versus avoid, reducing noise in the final evidence bundle. 
The effect is most visible on Yelp (large margin), where early GeoContext plus candidate semantics improves feasibility filtering and denoising before ranking.

\paragraph{Cold-Start.}
When item signals are sparse, the planner shifts toward candidate-side evidence via ItemSemantic/ItemProfile; when user history is scarce, it relies on stable cues distilled by LongTermPreference/ShortTermPreference, optionally refined by preference filters. 
This targeted routing avoids low-yield actions under sparsity and preserves whichever side carries the most information. 
Resulting margins are Amazon: CS-User $+17.75\%$, CS-Item $+23.20\%$; Goodreads: CS-User $+37.32\%$, CS-Item $+2.02\%$; Yelp: CS-User $+218.60\%$, CS-Item $+218.60\%$.

\begin{table*}[t]
  \centering
  \caption{Ablation study (Avg. HR@\{1,3,5\}) arranged by Domain $\rightarrow$ Scenario $\rightarrow$ Split. 
  \emph{Best Agentic RS (Tab.~\ref{tab:all_results_domain_first})} selects, \textbf{for each column}, the strongest agentic recommender baseline from Table~\ref{tab:all_results_domain_first} as a point-of-reference.}
  \label{tab:ablation_domain_first}
  \small
  \setlength{\tabcolsep}{3pt}
  \renewcommand{\arraystretch}{1.10}
  \begin{tabular*}{\textwidth}{@{\extracolsep{\fill}} l c c c c c c c c c c c c c c c}
    \toprule
    \multicolumn{1}{c}{} &
    \multicolumn{5}{c}{Amazon} &
    \multicolumn{5}{c}{Goodreads} &
    \multicolumn{5}{c}{Yelp} \\
    \cmidrule(lr){2-6}\cmidrule(lr){7-11}\cmidrule(lr){12-16}
    Method &
    Classic & \multicolumn{2}{c}{Cold-Start} & \multicolumn{2}{c}{Evo-Int} &
    Classic & \multicolumn{2}{c}{Cold-Start} & \multicolumn{2}{c}{Evo-Int} &
    Classic & \multicolumn{2}{c}{Cold-Start} & \multicolumn{2}{c}{Evo-Int} \\
    \cmidrule(lr){3-4}\cmidrule(lr){5-6}
    \cmidrule(lr){8-9}\cmidrule(lr){10-11}
    \cmidrule(lr){13-14}\cmidrule(lr){15-16}
    & & User & Item & Long & Short
      & & User & Item & Long & Short
      & & User & Item & Long & Short \\
    \midrule
    \textbf{ChainRec }
      & 72.3 & 70.3 & 60.0 & 68.7 & 74.3
      & 56.7 & 67.7 & 50.5 & 70.3 & 69.7
      & 28.3 & 13.7 & 13.7 & 15.7 & 14.0 \\
    w/o DPO
      & 69.6 & 70.3 & 57.0 & 67.3 & 69.3
      & 54.3 & 66.7 & 46.1 & 68.7 & 68.3
      & 27.7 & 12.7 & 13.7 & 13.7 & 13.0 \\
    Deepseek-R1 
      & 70.7 & 78.7 & 64.3 & 69.6 & 77.0
      & 57.7 & 69.3 & 48.1 & 70.0 & 65.7
      & 34.0 & 13.3 & 15.0 & 13.7 & 14.3 \\
    \midrule
    \emph{Best Agentic RS (Tab.~\ref{tab:all_results_domain_first})}
      & 63.0 & 59.7 & 48.7 & 65.0 & 71.3
      & 56.3 & 49.3 & 49.5 & 68.7 & 66.3
      & 7.6  & 4.3  & 4.3  & 10.3 & 7.7 \\
    \bottomrule
  \end{tabular*}
\end{table*}

\paragraph{Evolving-Interest.}
Under drift, the planner reweights ShortTermPreference vs.\ LongTermPreference based on the current state, then fuses evidence with ItemSemantic/ItemProfile (and, when helpful, PositivePreference/NegativePreference) for CandidateRank. 
This keeps recency-sensitive cues and long-horizon tastes in balance while maintaining a normalized, schema-aligned memory. 
Margins are Amazon: Long $+5.69\%$, Short $+4.21\%$; Goodreads: Long $+2.33\%$, Short $+5.13\%$; Yelp: Long $+52.43\%$, Short $+81.82\%$. 
Structured memory writes keep inputs consistent across scenarios and improve robustness over fixed CoT lists.

\subsection{Ablation Study}

To identify where gains come from while minimizing confounds, we run ablations along three axes. 
\textit{Optimization signal:} compare DPO with SFT-only to test whether improvements stem from better routing rather than shorter plans. 
\textit{Backbone strength:} replace the planner with Deepseek-R1 to measure the effect of a stronger model that tends to produce longer plans. 
\textit{Evidence type:} remove Yelp locality to quantify dependence on domain-specific cues. 
Unless noted, the protocol, prompts, and tool interface remain fixed. 
Tables~\ref{tab:avg_steps} and~\ref{tab:ablation_domain_first} report average planning steps and accuracy.

\begin{table}[H]
  \centering
  \caption{Average planning steps, computed as the mean across all domainsand scenarios.}

  \label{tab:avg_steps}
  \small
  \setlength{\tabcolsep}{10pt}
  \renewcommand{\arraystretch}{1.10}
  \begin{tabular}{lc}
    \toprule
    Method & Avg Steps \\
    \midrule
    \textit{Best agentic baseline} & -- \\
    \textbf{ChainRec}       & \textbf{5.05} \\
    \quad w/o DPO                  & 4.94 \\
    \quad Deepseek-R1              & 6.28 \\
    \bottomrule
  \end{tabular}
\end{table}

\begin{table}[t]
  \centering
  \caption{Ablation on Yelp \textsc{GeoContext}: average HR@\{1,3,5\}.}
  \label{tab:yelp_geo_ablation}
  \footnotesize
  \setlength{\tabcolsep}{2.2pt}
  \renewcommand{\arraystretch}{1.05}
  \resizebox{\columnwidth}{!}{%
  \begin{tabular}{lccccc}
    \toprule
    Method & Classic & CS-U & CS-I & EI-L & EI-S \\
    \midrule
    \emph{Best Agentic RS}   & 7.6  & 4.3 & 4.3 & 10.3 & 7.7 \\
    \textbf{ChainRec}        & \textbf{28.3} & \textbf{13.7} & \textbf{13.7} & \textbf{15.7} & \textbf{14.0} \\
    \quad w/o \textsc{Geo-Context} & 23.7 & 10.3 & 11.7 & 12.7 & 12.3 \\
    \midrule
    \textit{w/o vs ChainRec} 
                              & \textbf{-16.3\%} & \textbf{-24.8\%} & \textbf{-14.6\%} & \textbf{-19.1\%} & \textbf{-12.1\%} \\
    \textit{w/o vs Baseline} 
                              & \textbf{+211.8\%} & \textbf{+139.5\%} & \textbf{+172.1\%} & \textbf{+23.3\%} & \textbf{+59.7\%} \\
    \bottomrule
  \end{tabular}
  }
\end{table}

\paragraph{Ablation Study.}
With comparable plan length, DPO yields consistent gains: \textbf{ChainRec} and \textit{w/o DPO} average $5.05$ vs.\ $4.94$ steps (Table~\ref{tab:avg_steps}), yet the full model wins in key cells—Amazon–Evo-Short by $+5.0$ and Goodreads–CS-Item by $+4.7$ (Table~\ref{tab:ablation_domain_first})—indicating more effective state-matched routing under the same budget. 
Using Deepseek-R1 as a strong control, the longer-plan backbone scores higher on Amazon–CS-User ($+8.4$) and Yelp–Classic ($+5.7$), but lower on Goodreads–Evo-Short ($-4.0$) and Yelp–Evo-Long ($-2.0$). 
Deepseek-R1 averages $6.28$ steps versus $5.05$ for ChainRec (Table~\ref{tab:avg_steps}), showing that plan length alone does not guarantee better ranking. 
Taken together, the step counts for \textit{w/o DPO} and the full model are nearly identical while accuracy diverges most in sparsity/drift-sensitive cells; longer plans can help in some cases, but \emph{state-aware routing}—allocating steps to the most informative actions—ultimately governs quality under the fixed-candidate protocol.

Table~\ref{tab:yelp_geo_ablation} further highlights the role of domain-specific evidence.
Removing the Yelp locality cue causes the largest performance drops,
especially in User Cold-Start ($-24.8\%$) and Evolving-Interest (Long) ($-19.1\%$).
However, even without locality, ChainRec still surpasses the strongest agentic baseline
in the Classic setting by $+211.8\%$.
This suggests that locality cues are beneficial when available, but the overall robustness
primarily stems from the standardized tool capabilities and effective planner routing.

Overall, the ablations confirm that ChainRec’s gains come from improved tool composition
under a fixed interaction budget, rather than longer plans or reliance on any single cue.

\section{Conclusion}

In this work, we introduced \textsc{ChainRec}, an agentic recommendation framework designed for interactive settings where the system must actively gather evidence before ranking items. Unlike prior approaches that rely on fixed reasoning workflows, \textsc{ChainRec} allows the agent to flexibly decide what information to retrieve, how to combine it, and when to stop, depending on the recommendation scenario.

To support this adaptivity, we built a standardized library of reusable tools from expert reasoning trajectories, and trained a lightweight planner to compose these tools dynamically. This design makes the recommendation process more modular and controllable: tools provide consistent evidence summaries, while the planner focuses on selecting the most useful sequence under different conditions.

Experiments on AgentRecBench across Amazon, Goodreads, and Yelp demonstrate that \textsc{ChainRec} consistently improves ranking accuracy over strong baselines, with especially clear gains in cold-start and interest-shift scenarios. Overall, our results suggest that enabling agents to retrieve evidence strategically, rather than following scripted pipelines, is an effective direction for building more adaptive recommender systems.

Future work includes extending the framework to broader recommendation domains,
such as music, video, or news, to further evaluate its generality across diverse settings.

\bibliographystyle{ACM-Reference-Format}
\bibliography{sample-base}

\appendix

\section{Training Data Construction}\label{app:data}

\textbf{Motivation and scope.}
To comply with the evaluation protocol without touching test instances, we build a 4{,}500-task training corpus aligned with \emph{AgentRecBench} (the official testbed for the AgentSociety recommendation track). We use only interface-level aggregate statistics (e.g., candidate-set size, popularity buckets, recency windows). We do not access any instance-level records, texts, or labels from the test split, and we strictly de-duplicate user–item pairs against it. Evaluation follows the 20-candidate protocol (1 positive, 19 negatives).

\noindent\textbf{Scenario abstraction and over-generation.}
We follow the benchmark’s abstraction and visibility controls: Classic, Evolving-Interest (Long / Short with approximate 90-day / 7-day windows), and Cold-Start (user / item). For each \{domain (Amazon / Goodreads / Yelp) × scenario\} cell, we create a reference profile (aggregate statistics) and over-generate candidate tasks under the following constraints: (i) candidate-size alignment to the cell median within a bounded range; (ii) popularity alignment using the reference histogram of \texttt{rating\_count}; (iii) user/item bounds for history length and \texttt{rating\_count} derived from the median and 90th percentile with scenario-specific overrides; (iv) temporal alignment for Long/Short windows; and (v) cold-start thresholds via public quantile rules.

\begin{figure*}[t]
  \centering
  \includegraphics[width=\textwidth]{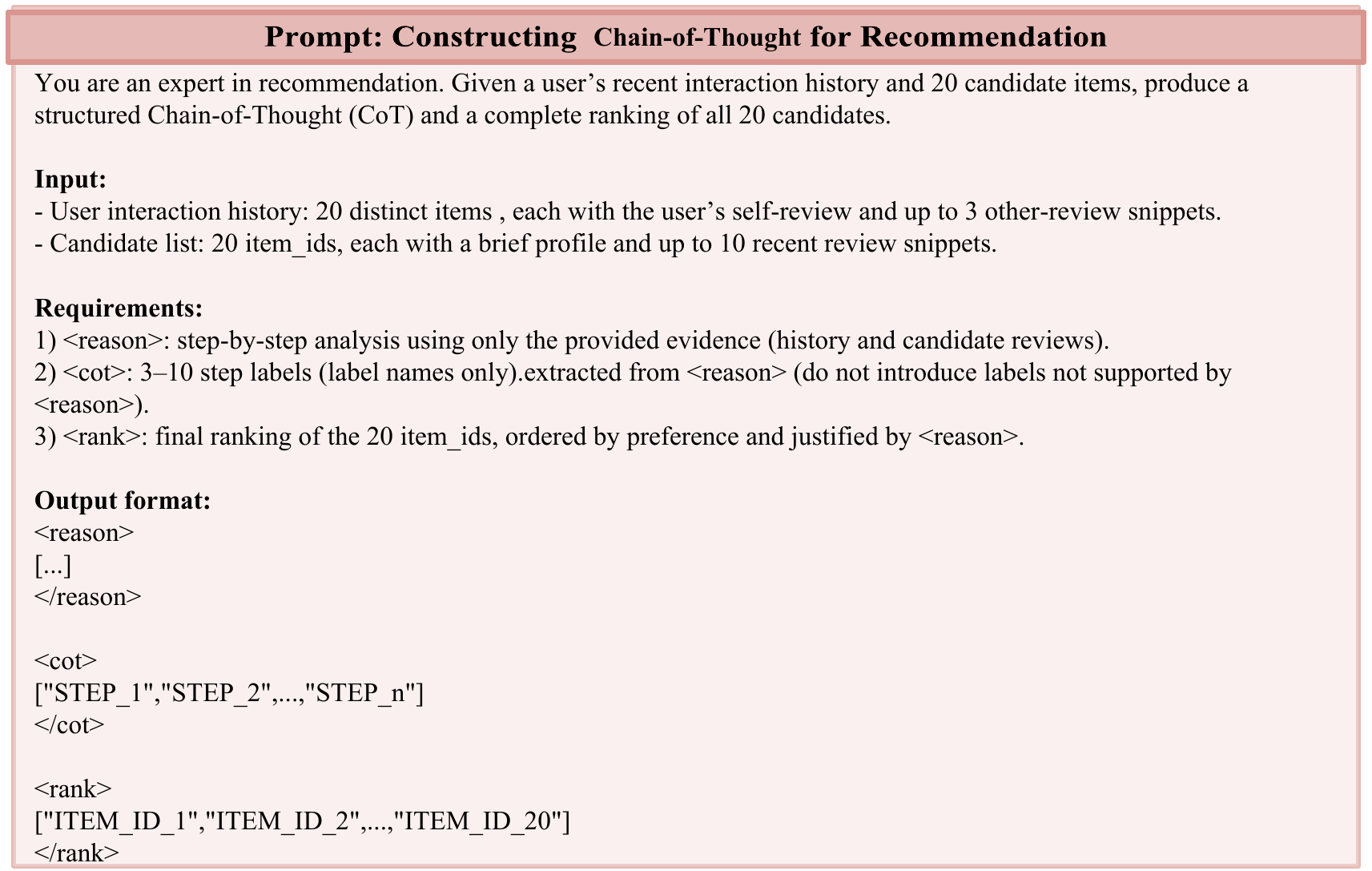}
  \caption{Prompt used to construct Chain-of-Thought (CoT) for recommendation. 
  The template specifies the input (user interaction history and 20 candidates), 
  requirements (step-by-step reasoning, CoT step labels, and a full ranking), 
  and the structured output format.}
  \label{fig:prompt-cot}
\end{figure*}

\noindent\textbf{Subset selection and distribution matching.}
From the over-complete pool, we apply forward greedy selection to obtain 4{,}500 tasks that best match the reference profiles. The objective combines: (a) a primary term—the L1 distance between candidate-popularity histograms; (b) secondary terms—the relative errors of mean candidate size, mean user-history length, and mean candidate popularity; (c) a Yelp-specific geo-coverage term; and (d) two joint statistics (candidate-size × popularity-bucket; history-length × recency-bucket) to avoid matching marginals only. We use multiple random restarts and keep the best subset; variance across seeds is small.

\section{Tool Construction}\label{app:tool-construction}

\noindent\textbf{CoT prompt.}
We elicit step-by-step CoT traces under the same interactive constraints as evaluation (20 fixed candidates, explicit step labels, and a valid final permutation). The exact template appears in Fig.~\ref{fig:prompt-cot}. Raw fields are redacted to interface-level placeholders, and input/output slots remain unchanged. To ensure the LLM has sufficient evidence to construct complete CoT and solve the task, for each episode we load the target user's most recent 10 interactions (purchases, reads, or consumptions, depending on the domain) and up to 3 reviews for each interacted item. We also load all 20 candidates specified by the protocol, each with metadata and up to 10 reviews. Review snippets are selected with simple recency and helpfulness heuristics and truncated to fit the token budget. This standardized context is fed into the prompting template and preserved in the normalized inputs and outputs used for capability mining.

\noindent\textbf{CoT embeddings.}
We extract step-level summaries from valid episodes and keep only samples where (i) the ground-truth item appears within the top-5 of the pre-ranking over the 20 candidates and (ii) the step field exists and is non-empty. Each CoT step is encoded as a 256-dimensional sentence vector using Zhipu AI’s embedding-3 model, and all vectors are L2-normalized before clustering.

\noindent\textbf{CoT k-means.}
We first normalize raw step texts into a small action vocabulary (op) with a few arguments (args), removing near-synonyms and noisy variants to reduce variance. We then encode steps as L2-normalized sentence vectors and cluster them with k-means. The number of clusters is selected by sweeping a small range and inspecting elbow/silhouette trends, preferring the smallest $k$ that preserves semantic coherence. We fix the random seed, use multiple initializations, and assign each step a cluster label. For each cluster we make a brief “cluster card” (size + representative bullets) to name patterns and map them into the Tool Agent Library. Quality checks include within-cluster consistency, coverage of major intents, and optional 2D projections; very small clusters may be merged into nearest neighbors. The final taxonomy is fixed for analysis. Because Yelp has fewer valid CoT episodes and some locality-specific patterns, we list its domain-specific clusters separately as [10].

\noindent\textbf{CoT cluster summary}
\begin{itemize}[leftmargin=*, itemsep=2pt, topsep=4pt]
\item \textbf{[1]} — Repeat purchase potential; Workflow synergy evaluation; Entertainment relevance check
\item \textbf{[2]} — Final relevance synthesis; Contextual elimination; Dynamic ranking synthesis
\item \textbf{[3]} — Genre alignment assessment; Superhero genre dominance analysis; Dark thematic affinity identification
\item \textbf{[4]} — Core interest matching; Positive preference matching; Negative preference filtering
\item \textbf{[5]} — Consolidated ranking; Final ranking synthesis; Final holistic prioritization
\item \textbf{[6]} — User behavior profiling; User profile synthesis; User preference profiling
\item \textbf{[7]} — Price; Item; Budget/value hints
\item \textbf{[8]} — Exploration–familiarity balance; Thematic novelty vs.\ familiarity; Pragmatic novelty consideration
\item \textbf{[9]} — Reader profile dissection; Reader demographic and habits; Reader role and preferences
\item \textbf{[10]} — Series/author continuity evaluation; Manga crossover potential; Series continuity assessment
\item \textbf{[11]} — Adjust for geographic accessibility; Location and transit (Yelp)
\end{itemize}

\noindent\textbf{From clusters to tools.}
We first derive a preliminary mapping by aligning each cluster’s dominant intent (reasoning role and typical inputs/outputs) with the tool interfaces. To reduce ambiguity, we then use an LLM to perform a secondary matching pass: for each cluster, the LLM reviews its centroid bullets and sampled members, predicts the most compatible tool role, and flags conflicts for manual spot checks. After this LLM‑aided validation, we map clusters to tools as follows:
[2],[5] $\rightarrow$ CandidateRank;
[1],[6],[9] $\rightarrow$ LongTermPreference;
[8] $\rightarrow$ ShortTermPreference;
[4] $\rightarrow$ PositivePreference / NegativePreference;
[3] $\rightarrow$ ItemSemantic;
[7] $\rightarrow$ ItemProfile;
[9],[10] $\rightarrow$ AuthorPreference;
[11] (Yelp) $\rightarrow$ GeoContext.
This yields a small, reusable action space for the planner.

\noindent\textbf{Tool summary.}
\begin{itemize}[leftmargin=*, itemsep=2pt, topsep=4pt]
\item \textbf{LongTermPreference} — Derives a long-horizon user profile as a stable anchor. \emph{Inputs:} Memory (long-window history) \emph{Outputs:} Long-term profile; concise summary.
\item \textbf{ShortTermPreference} — Captures recent intent and short-horizon shifts. \emph{Inputs:} Memory (recent interactions) \emph{Outputs:} Short-term signals; concise summary.
\item \textbf{PositivePreference} — Extracts positive cues (what to favor) from evidence. \emph{Inputs:} Memory (selected user evidence) \emph{Outputs:} Positive cues; concise summary.
\item \textbf{NegativePreference} — Identifies dislikes/constraints (what to avoid). \emph{Inputs:} Memory (selected user evidence) \emph{Outputs:} Negative cues; concise summary.
\item \textbf{ItemSemantic} — Summarizes item themes/semantics for alignment. \emph{Inputs:} Memory (candidate metadata/content). \emph{Outputs:} Per-item semantic tags/summaries.
\item \textbf{ItemProfile} — Compiles basic per-item profiles for matching/explanations. \emph{Inputs:} Memory (candidate metadata for the 20 items). \emph{Outputs:} Per-item profiles (structured).
\item \textbf{AuthorPreference} — Infers favored authors/series and affinity (domain-specific). \emph{Inputs:} Memory (user history; candidate author/series info). \emph{Outputs:} Author/series preferences; per-item affinity hints.
\item \textbf{GeoContext (Yelp)} — Assesses distance/accessibility and time feasibility. \emph{Inputs:} Memory (inferred user location; candidate locations/hours). \emph{Outputs:} Per-item geo scores; usability summary.
\item \textbf{CandidateRank} — Produces the final permutation with brief rationales. \emph{Inputs:} Memory (all collected evidence); fixed candidate set. \emph{Outputs:} Final ranked list; short explanations.
\end{itemize}

\section{Case Study}\label{app:case}
\begin{table*}[t]
  \centering
  \caption{Overall results with \textbf{Qwen} as the backbone LLM (Avg.\ HR@\{1,3,5\}).
  Organized as Domain $\rightarrow$ Scenario (Classic / Cold-Start / Evolving-Interest) $\rightarrow$ Split (User/Item or Long/Short).
  Baselines use the Qwen backbone; the last two rows add our method (placeholders) and its margin over the \emph{best agentic baseline} per column.}
  \label{tab:qwen_results_domain_first}
  \small
  \setlength{\tabcolsep}{1.5pt}
  \renewcommand{\arraystretch}{1.10}
  \begin{tabular*}{\textwidth}{@{\extracolsep{\fill}} l l ccccc ccccc ccccc}
    \toprule
    \multicolumn{2}{c}{} &
    \multicolumn{5}{c}{Amazon} &
    \multicolumn{5}{c}{G'reads} &
    \multicolumn{5}{c}{Yelp} \\
    \cmidrule(lr){3-7}\cmidrule(lr){8-12}\cmidrule(lr){13-17}
    \multicolumn{2}{c}{} &
    Classic & \multicolumn{2}{c}{Cold-Start} & \multicolumn{2}{c}{Evo-Int} &
    Classic & \multicolumn{2}{c}{Cold-Start} & \multicolumn{2}{c}{Evo-Int} &
    Classic & \multicolumn{2}{c}{Cold-Start} & \multicolumn{2}{c}{Evo-Int} \\
    \cmidrule(lr){3-3}\cmidrule(lr){4-5}\cmidrule(lr){6-7}
    \cmidrule(lr){8-8}\cmidrule(lr){9-10}\cmidrule(lr){11-12}
    \cmidrule(lr){13-13}\cmidrule(lr){14-15}\cmidrule(lr){16-17}
    Category & Method &
         & \makebox[1.8em][c]{User} & \makebox[1.8em][c]{Item}
         & \makebox[1.9em][c]{Long} & \makebox[1.9em][c]{Short} &
         & \makebox[1.8em][c]{User} & \makebox[1.8em][c]{Item}
         & \makebox[1.9em][c]{Long} & \makebox[1.9em][c]{Short} &
         & \makebox[1.8em][c]{User} & \makebox[1.8em][c]{Item}
         & \makebox[1.9em][c]{Long} & \makebox[1.9em][c]{Short} \\
    \midrule
    Traditional RS & MF
    & 15.0 & 15.0 & 15.0 & 38.4 & 49.5
    & 15.0 & 15.0 & 15.0 & 17.7 & 21.3
    & 15.0 & 15.0 & 15.0 & 42.0 & 65.9 \\
    DL-based RS & LightGCN
    & 15.0 & 15.0 & 15.0 & 32.3 & 59.1
    & 15.0 & 15.0 & 15.0 & 18.0 & 22.3
    & 15.0 & 15.0 & 15.0 & 31.6 & 68.9 \\
    \midrule
    \multirow{8}{*}{Agentic RS} & BaseAgent
    & 39.0 & 16.6 & 15.0 & 16.0 & 13.3
    & 15.7 & 16.0 & 21.2 & 21.3 & 25.0
    & 3.7  & 2.0  & 3.3  & 5.7  & 6.0 \\
    & CoTAgent
    & 39.0 & 15.7 & 17.3 & 14.0 & 15.0
    & 16.0 & 18.0 & 18.5 & 22.3 & 23.7
    & 5.0  & 2.3  & 2.3  & 3.3  & 4.3 \\
    & MemoryAgent
    & 37.3 & 16.3 & 12.0 & 10.3 & 14.0
    & 17.3 & 18.3 & 17.8 & 21.7 & 22.7
    & 5.3  & \textbf{3.3}  & \textbf{4.3}  & 4.0  & 3.7 \\
    & CoTMemAgent
    & 37.0 & 16.7 & 14.3 & 12.3 & 15.0
    & 16.7 & 17.7 & 20.2 & 23.3 & 22.0
    & 3.7  & 2.3  & 3.3  & 3.3  & 4.6 \\
    & Baseline666
    & \textbf{69.0} & \textbf{48.7} & 48.3 & 50.6 & 55.3
    & 45.0 & 36.7 & 44.1 & \textbf{63.0} & \textbf{55.7}
    & 7.0  & 1.3  & 2.3  & 0.0  & 0.0 \\
    & DummyAgent
    & 44.0 & 44.7 & 45.6 & \textbf{53.0} & 56.7
    & \textbf{46.0} & 39.3 & \textbf{47.8} & 54.0 & 54.0
    & 5.3  & 1.7  & 0.3  & 10.6  & 6.3 \\
    & RecHackers
    & 54.0 & 44.7 & \textbf{49.3} & 52.7 & \textbf{57.3}
    & 45.3 & 38.3 & 42.1 & 62.0 & 54.0
    & \textbf{7.7}  & 1.0  & 2.3  & \textbf{11.3}  & 5.3 \\
    & Agent4Rec
    & 23.3 & 25.7 & 48.3 & 26.3 & 37.7
    & 9.3  & \textbf{40.7} & 9.1  & 36.7 & 43.9
    & 5.7  & 2.3  & 0.7 & 8.6  & \textbf{9.0} \\
    \midrule
    \textbf{Ours} & \textbf{ChainRec (Qwen)}
    & \textbf{67.3} & \textbf{53.7} & \textbf{50.7} & \textbf{58.0} & \textbf{57.3}
    & \textbf{47.7} & \textbf{65.7} & \textbf{52.1} & \textbf{62.3} & \textbf{57.7}
    & \textbf{25.0} & \textbf{14.3} & \textbf{10.3} & \textbf{20.7} & \textbf{19.7} \\
     & \textit{$\Delta$ vs best (Qwen)}
  & {\footnotesize\itshape $-2.46\%$} & {\footnotesize\itshape $+10.27\%$} & {\footnotesize\itshape $+2.84\%$} & {\footnotesize\itshape $+9.43\%$} & {\footnotesize\itshape $+0.00\%$}
  & {\footnotesize\itshape $+3.70\%$} & {\footnotesize\itshape $+61.43\%$} & {\footnotesize\itshape $+8.99\%$} & {\footnotesize\itshape $-1.11\%$} & {\footnotesize\itshape $+3.59\%$}
  & {\footnotesize\itshape $+224.68\%$} & {\footnotesize\itshape $+333.33\%$} & {\footnotesize\itshape $+139.53\%$} & {\footnotesize\itshape $+83.19\%$} & {\footnotesize\itshape $+118.89\%$} \\

    \bottomrule
  \end{tabular*}
\end{table*}

\paragraph{Case Study: Yelp (Classic).}
\begin{itemize}
  \item \textbf{Tool Sequence.} PositivePreference $\rightarrow$ ItemSemantic $\rightarrow$ GeoContext $\rightarrow$ ItemProfile $\rightarrow$ CandidateRank.

  \item \textbf{Final Top-5 (titles).}
  \begin{itemize}
    \item 1) Divine Bovine Burgers
    \item 2) Frank's Sports Grill \& Bar
    \item 3) Breeze Patio Bar and Grill
    \item 4) The Bambi Bar
    \item 5) Murphy's Public House
  \end{itemize}

  \item \textbf{Top-1 vs Bottom-1 }
  \begin{itemize}
    \item \emph{Top-1} (Divine Bovine Burgers) --- Beer Bar/Bars/Nightlife/Restaurants/Burgers; rating $4.5$ ($n{=}294$).
    \begin{itemize}
      \item \textbf{ItemProfile:} Highly rated burger spot with excellent food reviews; early close (11:00--20:00 daily). Fit = high; popularity = high.
      \item \textbf{ItemSemantic:} ``Highly-rated burger bar with 4.5/294, closing at 8 PM''; tags = [Burgers, Beer Bar]; salient = [Excellent rating, High review count].
      \item \textbf{GeoContext:} distance\_km $= 0.44$; geo\_score $\approx 0.957$; global coverage $= 1.0$ (Geo usable).
      \item \textbf{PositivePreference.} Aggregated from observed interactions and cues:
  \begin{itemize}
    \item Friendly service (repeatedly mentioned as desirable).
    \item High food quality/freshness (delicious, fresh, top-quality items).
    \item Authentic ethnic cuisine (non-generic, authentic preparations).
    \item Restaurants/Dining categories often considered: Greek, Chinese, Mexican, seafood, burgers.
    \item Moderate interest in home services when customer support is good.
  \end{itemize}
      \item \textbf{CandidateRank:} score $\approx 0.85$; explanation: ``Highly-rated burger bar aligning with preferences for quality food and friendly service.''
    \end{itemize}

    \item \emph{Bottom-1} (Hooters) --- Bars/Nightlife/Chicken Wings/Sports Bars/Restaurants/American Traditional; rating $2.0$ ($n{=}45$).
    \begin{itemize}
      \item \textbf{ItemProfile:} Poorly rated location with repeated service complaints and wait-time issues; hours late (often 11:00--23:00, weekends until 00:00). Fit = low; popularity = medium.
      \item \textbf{ItemSemantic:} ``Sports bar with low 2.0/45 despite late hours''; tags = [Sports Bars, Chicken Wings]; salient = [Low rating, Late hours].
      \item \textbf{GeoContext:} distance\_km $= 1.593$; geo\_score $\approx 0.853$ (not far, but Geo is auxiliary).
      \item \textbf{PositivePreference.} Aggregated from observed interactions and cues:
  \begin{itemize}
    \item Friendly service (repeatedly mentioned as desirable).
    \item High food quality/freshness (delicious, fresh, top-quality items).
    \item Authentic ethnic cuisine (non-generic, authentic preparations).
    \item Restaurants/Dining categories often considered: Greek, Chinese, Mexican, seafood, burgers.
    \item Moderate interest in home services when customer support is good.
  \end{itemize}
      \item \textbf{CandidateRank:} score $\approx 0.00$; explanation: ``Low rating and service issues conflict with stated preferences.''
    \end{itemize}
  \end{itemize}

  \item \textbf{Takeaway.} The top item aligns with preference signals (friendly service, high food quality) and is supported by semantic/profile evidence, with helpful geo proximity. The bottom item, despite late hours and reasonable proximity, is penalized by low rating and service complaints, which conflict with the stated preferences. Geo acts as an auxiliary factor; the main judgment comes from preference–semantic–profile alignment and reliability.
\end{itemize}


\section{Results with Qwen Backbone}\label{app:altllm-qwen}

Replacing the backbone LLM with Qwen2.5-72B preserves the main trend observed in the main results. ChainRec (Qwen2.5-72B) exceeds the strongest agentic baseline in 13 of 15 columns and ties in one, with the largest margins appearing in Goodreads--Cold-Start (User) and across all Yelp settings. These gains align with our design goal: standardized capabilities plus learned routing allow the planner to pivot toward candidate-side evidence when user signals are sparse, and to fuse short-/long-horizon cues under interest drift, independent of the specific backbone.

Two cells show small deficits against the best baseline (Amazon--Classic and Goodreads--Evo-Long). Both correspond to stable, high-signal regimes where long prompts or fixed call orders already perform competitively; ChainRec remains close while retaining the flexibility to adapt across the harder cells.

The overall pattern mirrors the results with other backbones: improvements concentrate in cold-start and evolving-interest scenarios, and hold across domains including Yelp. This suggests that the contribution comes primarily from the capability–policy separation and state-aware routing, rather than from idiosyncrasies of any single LLM backbone.

\end{document}